\documentclass[aps,pra,showpacs,amsfonts,notitlepage,amssymb,amsmath,nofootinbib,superscriptaddress,twocolumn,longbibliography]{revtex4-1}
\usepackage[utf8]{inputenc}
\usepackage[english]{babel}
\usepackage[titletoc,toc,title]{appendix}
\usepackage{amsmath}
\usepackage{amsfonts}
\usepackage{amssymb}
\usepackage[colorlinks=true,linkcolor=blue,citecolor=blue,urlcolor=blue]{hyperref}
\usepackage{graphicx}
\usepackage{enumerate}
\usepackage{csquotes}
\usepackage[caption=false]{subfig}
\usepackage{dsfont}
\usepackage{color}
\usepackage{bbold}
\usepackage{mathtools}
\usepackage{tensor}
\usepackage{gensymb}

\begin{document}

\title{Correlations on weakly time-dependent transcritical white-hole flows}

\author{Johan Fourdrinoy}
\affiliation{Institut Pprime, CNRS - Université de Poitiers - ISAE-ENSMA. TSA 51124, 86073 Poitiers Cedex 9, France}
\author{Scott Robertson}
\affiliation{Laboratoire Charles Fabry, Institut d'Optique Graduate School, CNRS, Universit\'{e} Paris-Saclay, 91127 Palaiseau, France}
\affiliation{Universit\'e Paris-Saclay, CNRS/IN2P3, IJCLab, 91405 Orsay, France}
\author{Nicolas James}
\affiliation{Laboratoire de Mathématiques et Applications, Université de Poitiers, 11 Boulevard Marie et Pierre Curie, TSA 61125,  86073 Poitiers Cedex 9, France}
\author{Alessandro Fabbri}
\affiliation{Departamento de F\'{i}sica Te\'{o}rica and IFIC, Centro Mixto Universidad de Valencia--CSIC, C. Dr. Moliner 50, 46100 Burjassot, Spain}
\affiliation{Universit\'e Paris-Saclay, CNRS/IN2P3, IJCLab, 91405 Orsay, France}
\author{Germain Rousseaux}
\affiliation{Institut Pprime, CNRS - Université de Poitiers - ISAE-ENSMA. TSA 51124, 86073 Poitiers Cedex 9, France}

\begin{abstract}
We report on observations made on a run of transcritical flows over an obstacle in a narrow channel.  Downstream from the obstacle, the flows decelerate from supercritical to subcritical, typically with an undulation on the subcritical side (known in hydrodynamics as an undular hydraulic jump).  In the Analogue Gravity context, this transition corresponds to a white-hole horizon.  Free surface deformations are analyzed, mainly via the two-point correlation function which shows the presence of a checkerboard pattern in the vicinity of the undulation.  In non-gated flows where the white-hole horizon occurs far downstream from the obstacle, this checkerboard pattern is shown to be due to low-frequency fluctuations associated with slow longitudinal movement of the undulation.  
It can thus be considered as an artifact due to a time-varying background.
In gated flows, however, the undulation is typically ``attached'' to the obstacle, and the fluctuations associated with its movement are strongly suppressed.
In this case, the observed correlation pattern is likely due to a stochastic ensemble of surface waves, scattering on a background that is essentially stationary.
\end{abstract}

\maketitle

\section{Introduction}

Correlations provide valuable insight into the behavior of fluctuations.
They are the observable of choice in (quantum) field theory, where fluctuations are intrinsic and correlation functions indispensable.
In fluid mechanics they have found utility in the theory of turbulence~\cite{Wallace-2014}, though they are also useful in the analysis of wave-current interaction when the waves present are due to random noise~\cite{Euve-et-al-2016}.
That is, if we decompose the flow into a ``background'' state (in some sense an average value, typically a time-average) and fluctuations around this background, the fluctuations can be considered as a statistical ensemble of waves which interact with each other, or (if their amplitudes are sufficiently small) with the background mean flow alone. 

The latter case corresponds to the regime of Analogue Gravity~\cite{Unruh-1981,Weinfurtner-et-al-2011,Euve-et-al-2016}, which aims to simulate gravitational phenomena using condensed matter experiments~\cite{LivingReview,barcelo2019analogue}. 
To this end, the background is identified with an effective spacetime metric for the fluctuations, which play the role of test waves propagating in the effective spacetime.
As a realization of field theory in curved spacetime, Analogue Gravity finds great utility in correlation functions~\cite{Balbinot-et-al-2008,Carusotto-et-al-2008}.
In particular, they capture the pair-wise nature of the analogue of Hawking radiation from an effective horizon.
At an analogue white-hole~\footnote{A {\it white hole}, or sometimes {\it white fountain}, is the time-reversed version of a black hole, into which nothing can enter and from which everything is ejected.} horizon where the flow passes from supercritical to subcritical, these correlations are particularly involved: the expected two-point correlation function exhibits a checkerboard pattern~\cite{Mayoral-et-al-2011}, due in that case to correlations between two short-wavelength dispersive modes of opposite energy.
(We shall have more to say about analogue Hawking radiation in the discussion section at the end of this paper.)

That being said, there can be subtleties in how the background (or, in the context of Analogue Gravity, the effective metric) is to be defined, which in turn affects the identification of the fluctuations themselves.
For example, in a quantum system where averages are taken over an ensemble of experimental realizations, classical fluctuations of the mean field may pollute the quantum signal (see~\cite{steinhauer2014observation,Wang-et-al-2017,Kolobov-et-al-2021} for such an example involving density fluctuations in a Bose-Einstein condensate).
In a purely classical context, such as a steady water flow which is expected to be statistically stationary, the background can be identified as the constant component of the field ({\it i.e.}, its time-average) while the fluctuations capture the entirety of the time-dependence of the field. 
However, this approach can appear too crude if there is some weak time-dependence of the background.  Indeed, since we define the background via averaging in time, there is some ambiguity in the range or interval duration over which such averaging is appropriate.  By adopting a certain interval duration which is significantly shorter than the full duration of the recording, we allow for a degree of time-dependence in the background, which will inevitably show some fluctuations from one subinterval to the next.  This subinterval averaging effectively divides all the time-dependence into ``slow'' fluctuations (which can be associated with some movement of the background) and ``fast'' fluctuations (which occur on top of the background).  In this way, the motion of the background can to some extent be separated from the field evolution due to the presence of a stochastic ensemble of surface waves.

In this paper, 
we consider free surface deformations on a particular class of 1D water flows, where we see a noticeable degree of weak time-dependence of the background.
The flows are transcritical (in contrast to previous experimental Analogue Gravity works in water flows which have tended to be purely subcritical~\cite{Weinfurtner-et-al-2011,Euve-et-al-2016}). 
The effective metric valid in the hydrodynamic (long-wavelength) limit contains a white-hole horizon~\footnote{The occurrence of such a white-hole horizon, where the flow passes from supercritical to subcritical, is highly non-trivial in water flows.  It cannot be stably realised using the geometry of a single obstacle alone (see Fig.~1(d) of~\cite{Pratt-1984} and the discussion around it).  We thus rely here on dissipative effects to induce the transition via an undular hydraulic jump, or on the backwater effect induced by lowering the gate which yields a short supercritical region on top of the obstacle.}, and it is downstream from this white-hole horizon that the time-dependence of the background is apparent.
The flow can be regulated by the partial lowering of a gate at the downstream end of the channel.  For flows that are unobstructed at the downstream end, the white-hole horizon is generated by an undular hydraulic jump~\cite{Chanson-Montes-1995} some distance downstream from the obstacle, and the origin of the correlation pattern is shown to be a slow drift in the longitudinal position of the jump.  The associated frequencies are so low that it makes sense to separate this drift from the familiar surface waves and to treat it instead as a slow movement of the background; when this is done, the remaining ``fast'' fluctuations have a correlation pattern in which the checkerboard is strongly suppressed. 
Contrastingly, in flows with a partially closed gate at the downstream end (inducing a ``backwater effect''~\cite{ChansonBook} that affects the upstream part of the flow), the undulation is typically seen to be ``attached'' to the obstacle, and while there is still a noticeable degree of time-dependence in the downstream region, we observe no clear contribution to the two-point correlations associated with a slow drift of the background.  We show that, in this case, the low-frequency contribution to the full two-point function is much less significant than in non-gated flows. 
What this means for the interpretation of the observed correlations is discussed; we believe they stem from the ``true'' scattering of surface waves.

The paper is organized as follows.
Section~\ref{sec:Setup} provides a description of the experiment: the details of the water channel, the cameras used to record the free surface, and the numerical post-processing performed on the data.
In Section~\ref{sec:Observations}, we describe our observations, paying particular attention to slow movement on long time scales and the associated patterns in the two-point correlation function.  This falls into two parts, with subsections~\ref{sec:nongated} and~\ref{sec:gated} dealing with non-gated and gated flows, respectively.
We summarize our findings in Section~\ref{sec:Discussion} and discuss some of the implications.
More information and technical details are given in the ``Materials and Methods'' section.

\section{Experimental setup}
\label{sec:Setup}

\begin{figure}
\centering
\includegraphics[width=7.8cm]{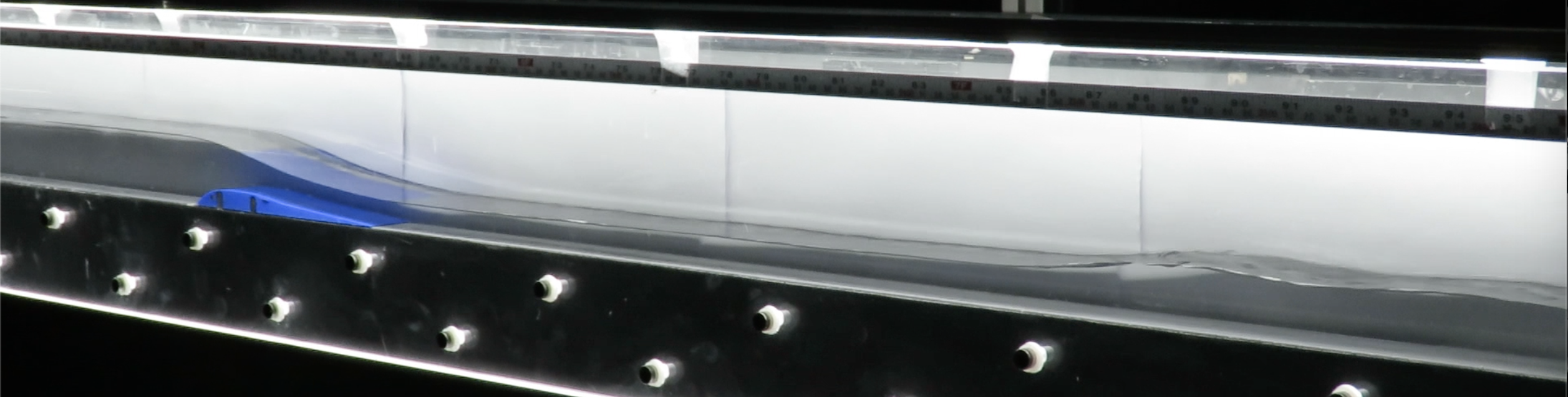}
\caption{Photograph of the channel, with an ABS obstacle (in blue) placed on the bottom.  The variation of the water height can be clearly seen, including the undular hydraulic jump on the far downstream (right) side of the flow.}
\label{fig:figure1}
\end{figure}

We used an 
open flow channel 
(reference H23 from Prodidac), of which a photograph is shown in Figure~\ref{fig:figure1}.  The channel walls are 
made of transparent plexiglass with anodized aluminum support, with 
dimensions ($L=2.5~{\rm m}$) $\times$ ($Z=12~{\rm cm}$) $\times$ ($W=5.3~{\rm cm}$). 
The maximum flow rate is $35~{\rm L}/{\rm min}$, provided by a volumetric hydraulic power bench. 
The flow rate is measured with a flowmeter Vortex F 20 (DN20) from Bamo Mesures, with a range from $5$ to $85~{\rm L}/{\rm min}$.  
Obstacles to be placed on the channel floor are made by an Ultimaker 5S 3D printing machine with the Cura software, 
using either PLA (black) or ABS (blue) filaments of diameter $2.85~{\rm mm}$. These obstacles have notches 
measuring $6~{\rm mm}$ in width and $3.5~{\rm mm}$ in depth, with which they are fixed on both sides of the channel. 
The presence of the obstacle forces a modulation of the flow due to the variation in the geometry of the flume.  As well as the obstacle, there is a gate at the downstream end of the flume that can be lowered to control the flow.

An overhead LED lighting system illuminates the free surface of the flow and allows the side visualization of the meniscus, which is recorded by two grayscale (256) Point Grey cameras with CMOS technology.  The images from the two cameras are combined by a Matlab algorithm. 
They record a total length of $2.05~\rm{m}$ with spatial resolution $\delta x = 0.5~{\rm mm}$, and (for the experiments here considered) a total duration of around 5 minutes at an acquisition rate of $f_{\rm{ac}}=32~\rm{fps}$. 
No wave maker is used; instead, the fluctuations of the free surface are provided by the noise inherent to the system (turbulence, mechanical vibrations, {\it etc.}).

The Matlab script processes the interface with a subpixel detection method applied to the side meniscus (see Refs.~\cite{Weinfurtner-et-al-2011,Faltot-et-al-2014,Euve-et-al-2016,euve2017classical,euve2020scattering,Mordant-et-al-2020} for details). 
The meniscus shows a maximum intensity on the reconstituted image. A first calculation makes it possible to detect this maximum for each position 
$x$ and each time 
$t$. After a first detection of maxima, the aberrant points (due to image problems like drops, blurs, {\it etc.}) are replaced by an average value of their neighbors, 
and the maximum brightness is sought again around the positions previously found. This two-step detection has a precision of one pixel. It is then followed by a subpixel approach: around its maximum value, the luminosity is assumed to decrease in the vertical direction according to a Gaussian (normal) distribution.  By fitting the observed luminosity to a Gaussian over five neighboring points, it is possible to find the position of the meniscus to within a fraction of the pixel size 
$\delta x$.

\section{Observations}
\label{sec:Observations}

A series of transcritical flows were realised, and recorded for a total duration of $\sim 340$ seconds. 
By ``transcritical'', we mean that the Froude number ${\rm Fr} = v/c$ (where $v$ is the flow velocity and $c$ the speed of surface waves with respect to the flow) crosses $1$.  In our narrow flume, we find that ${\rm Fr}$ tends to always be smaller than $1$ at the ends, and therefore that any transitition from subcritical to supercritical ({\it i.e.}, from ${\rm Fr} < 1$ to ${\rm Fr} > 1$) is followed somewhere by the opposite transition; that is, a black-hole horizon is typically followed by a white-hole horizon.

We present here two main cases, distinguished by the presence or absence of a gate at the downstream end of the channel.  We observe that this feature seems to determine whether the white-hole horizon occurs close to the obstacle, or a significant distance downstream from it.
We also observe that this property has implications for the content of the fluctuations of the free surface around its mean profile. 

Here we describe the most important observations, for non-gated flows in \S\ref{sec:nongated} and for gated flows in \S\ref{sec:gated}.  Technical details on how the key quantities are defined and extracted, as well as some supplementary observations, can be found in the Supplemental Material.

\subsection{Non-gated flows
\label{sec:nongated}}

In the absence of a gate at the downstream end, the flow typically remains supercritical for a significant distance downstream from the obstacle, then returns to subcriticality via an undular hydraulic jump: the water height increases quite abruptly, and is followed by an undulation of gradually decreasing amplitude~\cite{Chanson-Montes-1995}.  The jump occurs far in the downstream region, and is thus unlikely to be driven by the geometry of the obstacle.  This makes it quite different from the flows that are typically considered in theoretical Analogue Gravity works.  Instead, the form of the jump (and the white-hole horizon associated with it) is likely to be controlled by dissipative processes induced by turbulence.\footnote{Note that a dissipative term associated with turbulence is a required ingredient in theoretical treatments of the undular hydraulic jump; see, {\it e.g.}, Refs.~\cite{Steinrueck-et-al-2003,Jurisits-et-al-2007,Murschenhofer-Schneider-2019}.}  Such a hydraulic jump was not observed in our previous experimental work involving transcritical flows, in which a wider channel was used~\cite{euve2020scattering}.  We thus conjecture that the narrowing of the channel with respect to previous works (5.3 cm here compared to 39 cm in~\cite{Euve-et-al-2016,euve2020scattering}) increases the effective dissipation, leading generically to the presence of a jump whenever the flow becomes supercritical, and whenever the downstream end of the flow is free so that there is no backwater effect~\cite{ChansonBook} due to the gate.

\begin{figure}
\includegraphics[width=\columnwidth]{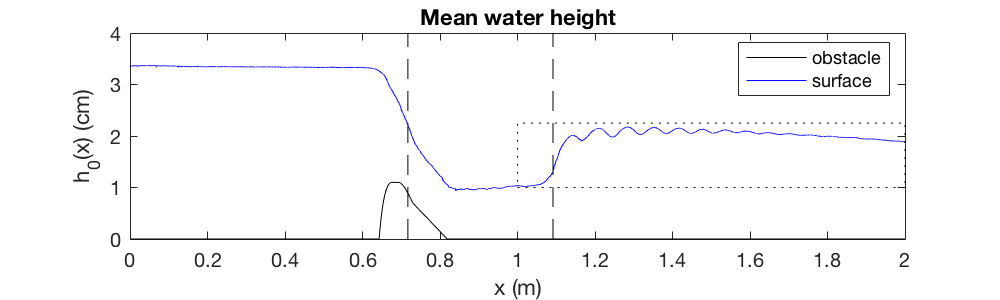} \\
\includegraphics[width=\columnwidth]{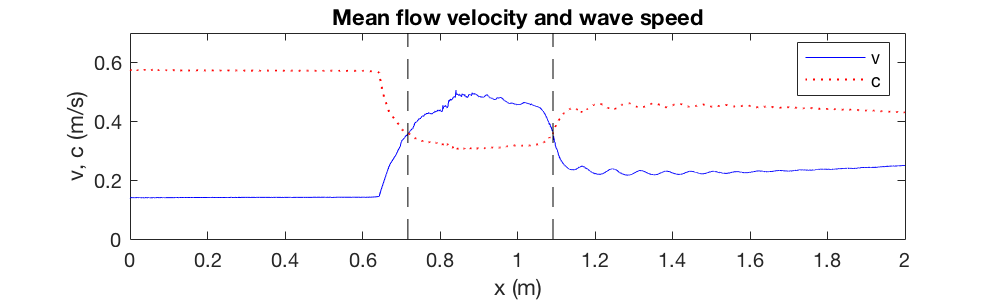} \\
\includegraphics[width=\columnwidth]{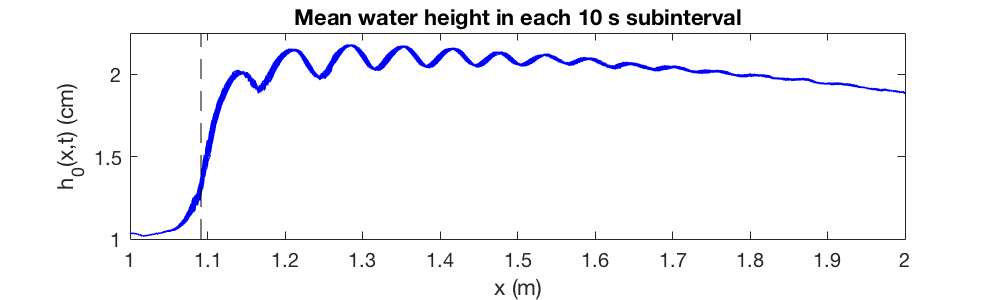} \\
\includegraphics[width=\columnwidth]{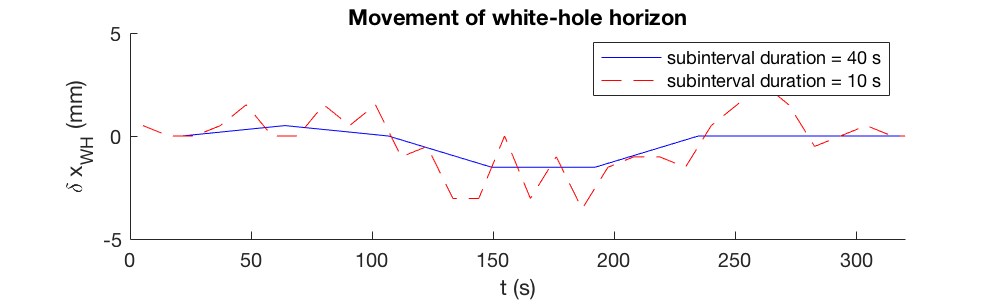}
\caption{We show here a series of plots characterizing a flow which is free (non-gated) at the downstream end. 
These are: 
(1) The obstacle height and the mean water height, averaged over the entire recording.  In vertical dashed lines are shown the positions of the black- and white-hole horizons.  
(2) The flow velocity $v(x)$ and wave speed $c(x)$.
(3) A zoom on the region shown by a dotted rectangle in panel (1), but now with the mean water height calculated separately in a series of $\sim 10 \, {\rm s}$ subintervals, all shown simultaneously.  The thickness of the curve thus indicates the degree of variation in time.  
(4) The position of the white-hole horizon, calculated in subintervals of duration $\sim 40 \, {\rm s}$ (in sold blue) and $\sim 10 \, {\rm s}$ (in dashed red).  
The flow rate is $Q=15\,{\rm L}/{\rm min}$, or (reducing to two dimensions) $q = Q/W = 4.7 \times 10^{-3} {\rm m}^{2}/{\rm s}$ (where $W = 5.3\,{\rm cm}$ is the channel width).
}
\label{fig:nogate}
\end{figure}

\begin{figure}
\includegraphics[width=\columnwidth]{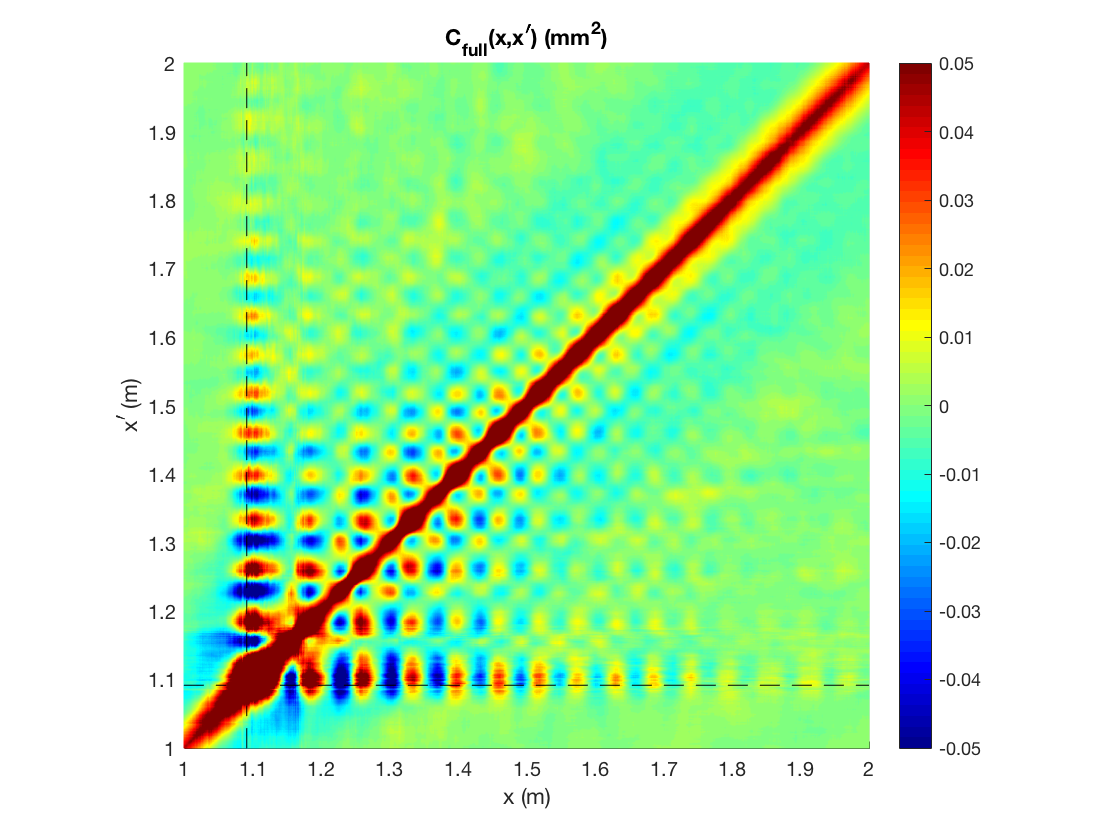}\\
\includegraphics[width=\columnwidth]{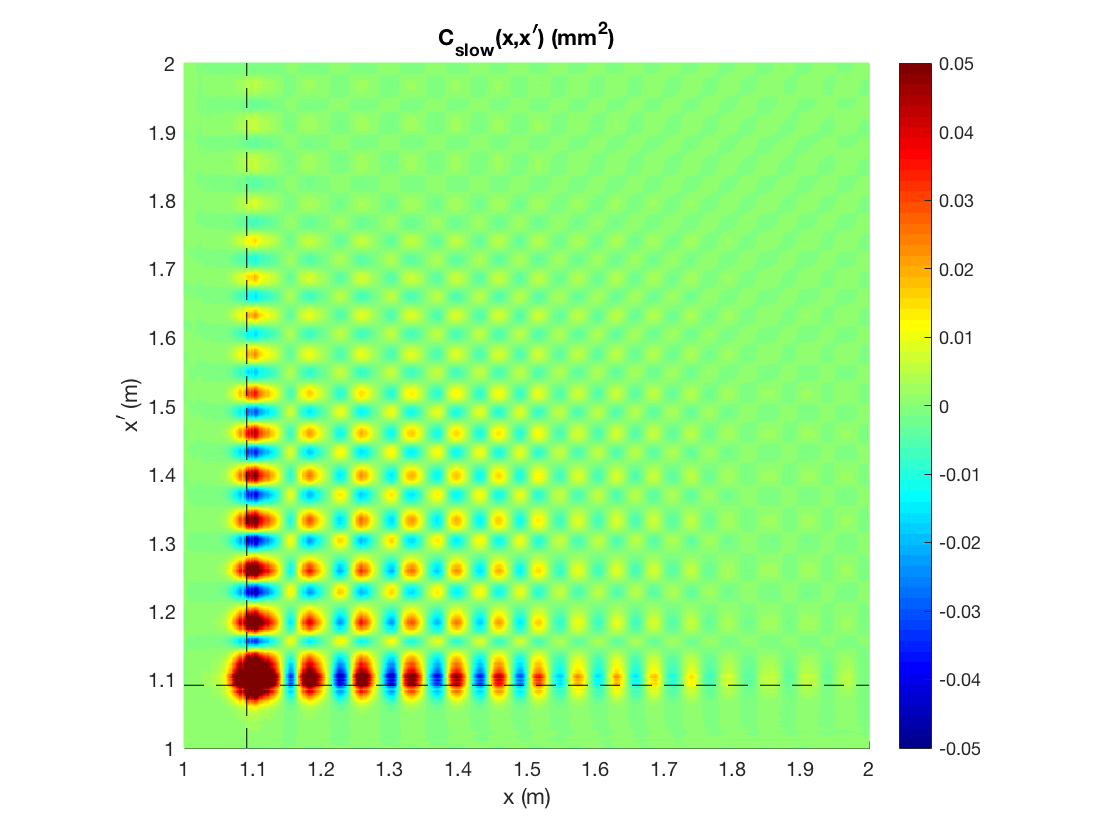}\\
\includegraphics[width=\columnwidth]{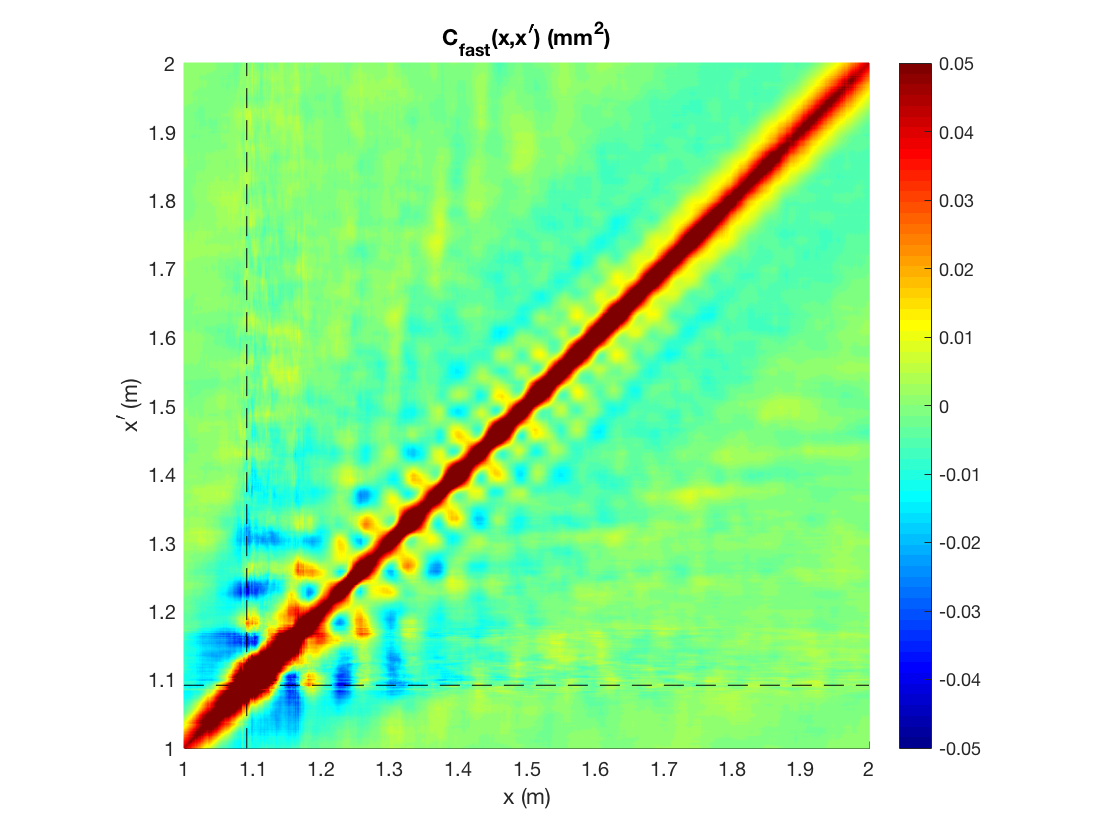}
\caption{The two-point correlation function of free surface fluctuations in the downstream region, for the same non-gated flow of Fig.~\ref{fig:nogate}. 
The dashed lines indicate the average position of the white-hole horizon. 
The three correlation functions correspond to different definitions of the fluctuation $\delta h$: the full fluctuations (where the background is the time-average over the entire recording), the ``slow'' fluctuations associated with movement of the background (found by averaging over 10-second subintervals), and the ``fast'' fluctuations on top of this slowly-varying background.  
}
\label{fig:nogate_checkerboard}
\end{figure}

\begin{figure}
\includegraphics[width=\columnwidth]{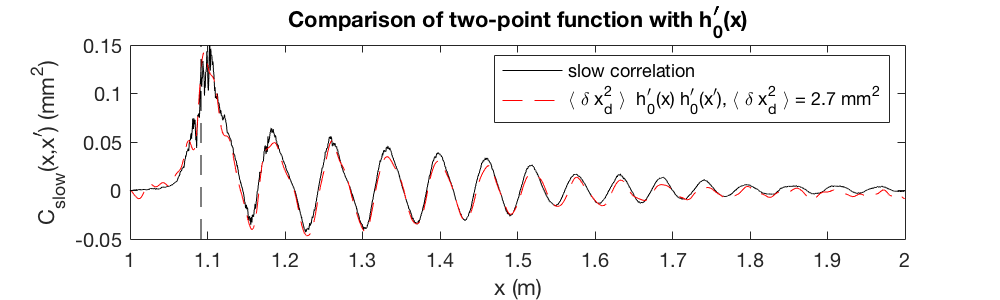} 
\caption{In black is shown a cross-section of the ``slow'' correlation of Fig.~\ref{fig:nogate_checkerboard}, with $x^{\prime} \approx 1.1 \, {\rm m}$ fixed close to the point where $C_{\rm slow}(x,x^{\prime})$ reaches its maximum value.  In red is shown the spatial derivative of the mean water height $h_{0}(x)$ (averaged over the entire recording), multiplied by a prefactor found by fitting it to the black curve.  Through prediction~(\ref{eq:corr_pred}), this prefactor gives the variance $\left\langle \delta x_{d}^{2} \right\rangle = 2.7 \, {\rm mm}^{2}$ of the longitudinal position of the undular jump.}
\label{fig:nogate_cslow}
\end{figure}

In Figure~\ref{fig:nogate} are shown four plots related to a non-gated flow (over an obstacle which is a scaled-down version of that used in Ref.~\cite{Weinfurtner-et-al-2011}). 
In the first panel, the mean water height $h_{0}(x)$, averaged over the entire 
duration of the recording, is shown, with the expected positions of the black- and white-hole horizons indicated by dashed lines.
The second shows the profiles of the flow velocity $v(x)$ and the wave speed $c(x)$, as calculated using Eqs.~(\ref{eq:v_and_c}) in the Supplemental Material.
In the third panel are shown the background profiles $h_{0}(x,t)$ averaged separately in each of 32 equal subintervals of $\sim 10\,{\rm s}$ duration each, having zoomed in on the region of interest 
where the non-stationarity of the background is most prominent.  The various $h_{0}(x,t)$ are shown simultaneously, so the thickness of the curve illustrates the degree of variation of the background over the entire recording.  
Notice that, while the upstream side of the flow is relatively stable~\footnote{The stability of the upstream part of the flow, around the analogue black-hole horizon, allows us to conclude that the time-dependence we see further downstream is not due to the pump -- and therefore the flow -- having yet to stabilize.  It seems instead to be an inherent feature of the system.}, there is a significant degree of variation of the background on the downstream side, and the beginning of the growth of this variation appears to coincide with the position of the undular hydraulic jump.
To get an idea of how the background actually varies in time from one subinterval to the next, the measured position of the white-hole horizon is shown in the fourth panel, suggesting a slow oscillatory nature.

In Figure~\ref{fig:nogate_checkerboard} we turn our attention to the two-point correlation function for free surface deformations: $C(x,x^{\prime}) = \left\langle \delta h(x) \, \delta h(x^{\prime}) \right\rangle$.  We focus on the region of interest in the vicinity of the undular hydraulic jump, {\it i.e.}, $x > 1\,{\rm m}$ (marked by the dotted rectangle in the top panel of Fig.~\ref{fig:nogate}).  
In the first panel, the background to be subtracted is taken as the time-average over the entire duration of the recording, so that all time-dependence of the free surface is contained in the fluctuations.  In the second and third panels, the background is defined and subtracted separately in 32 equal subintervals of duration $\sim 10 \, {\rm s}$. 
This procedure effectively defines a time-dependent background and divides the data into ``slow'' and ``fast'' fluctuations: those associated with movement of the background from one subinterval to the next (fluctuations of the background), and those occurring within each subinterval (fluctuations on top of the background). 
The most prominent feature, clearly visible in both the full and ``slow'' correlation functions, 
is a checkerboard pattern in the vicinity of the undulation.  The ``fast'' correlation function is essentially just the autocorrelation along $x=x^{\prime}$, indicating only that fluctuations are present but not necessarily correlated.

The observations described above lead us to make the following postulate on  the nature of the ``slow'' fluctuations giving rise to the checkerboard pattern of Fig.~\ref{fig:nogate_checkerboard}: namely, that they are 
mainly due to small longitudinal shifts in the position of the jump and the undulation.~\footnote{In a different context (namely a 1D Bose gas on which phonons propagate), the randomness in the position of the white-hole horizon from realization to realization might lead to a similar checkerboard-like correlation pattern due to the varying phase of the undulation~\cite{Wang-et-al-2017,Kolobov-et-al-2021}.}
To this end, we adopt the following ansatz for the profile of the undular hydraulic jump at any given time: $h_{0}(x,t) \approx h_{0}\left(x - \delta x_{d}(t)\right)$, where $\delta x_{d}(t)$ is the instantaneous shift in the position of the jump with respect to its mean position.  Assuming that this shift is always sufficiently small -- in particular, that it remains at all times much smaller than the wavelength of the undulation -- we can make a first-order Taylor expansion of this expression to get the instantaneous fluctuation of the background profile: $\delta h_{0}(x,t) \approx - \delta x_{d}(t) \, h_{0}^{\prime}(x)$.  Then the equal-time two-point correlation function associated with these fluctuations is
\begin{equation}
\left\langle \delta h_{0}(x,t) \, \delta h_{0}(x^{\prime},t) \right\rangle = \left\langle \delta x_{d}^{2}(t) \right\rangle \, h_{0}^{\prime}(x) h_{0}^{\prime}(x^{\prime}) \,.
\label{eq:corr_pred}
\end{equation}
This prediction is borne out by Fig.~\ref{fig:nogate_cslow}, where the two-point correlation function associated with ``slow'' background fluctuations is shown along a line of fixed $x^{\prime}$ (chosen to be close to the maximum of the two-point function).  This is plotted alongside the derivative of the mean water height, $h_{0}^{\prime}(x)$, after multiplication by a fitting parameter.
The two curves agree reasonably well, corroborating our claim that the checkerboard pattern is due to a slow longitudinal drift of the undulation.
According to prediction~(\ref{eq:corr_pred}), the value of the fitting parameter multiplying $h_{0}^{\prime}(x)$ is to be equated with $\left\langle \delta x_{d}^{2} \right\rangle \, h_{0}^{\prime}\left(x^{\prime}\right)$; the fitting procedure thus yields an estimate for the variance of $\delta x_{d}$, which is given in the legend of each plot and whose associated r.m.s value is $1.6 \, {\rm mm}$.  This in turn seems to be consistent with the spread of the position of the white-hole horizon shown in the fourth panel of Fig.~\ref{fig:nogate}.

\subsection{Gated flows
\label{sec:gated}}

\begin{figure}
\includegraphics[width=\columnwidth]{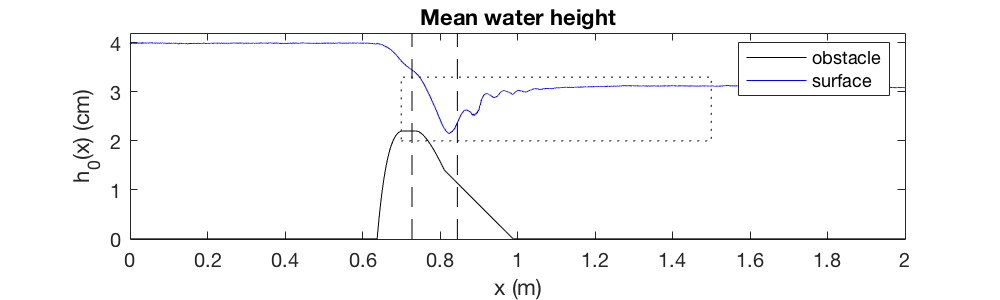} \\
\includegraphics[width=\columnwidth]{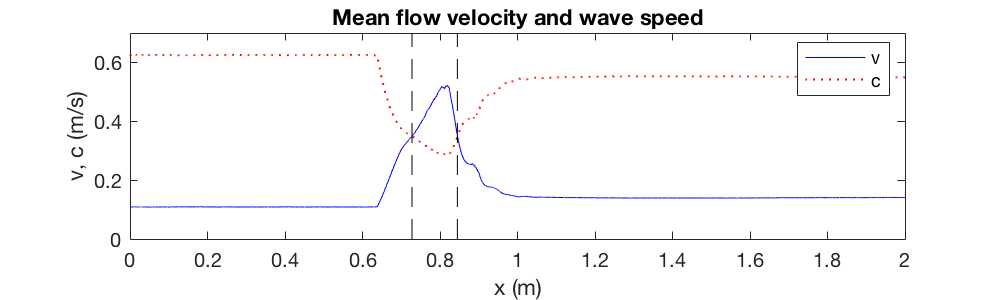} \\
\includegraphics[width=\columnwidth]{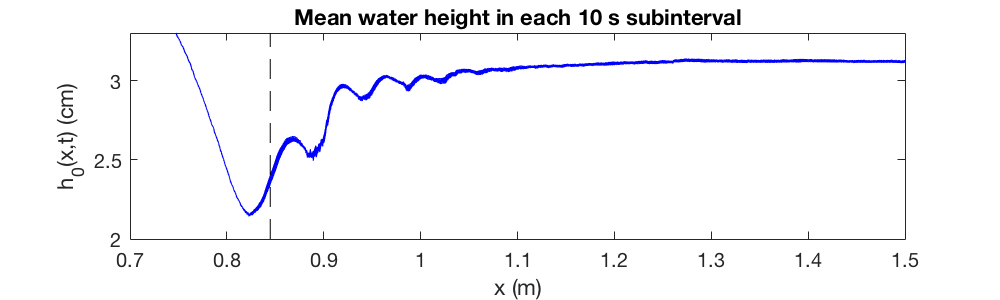} \\
\includegraphics[width=\columnwidth]{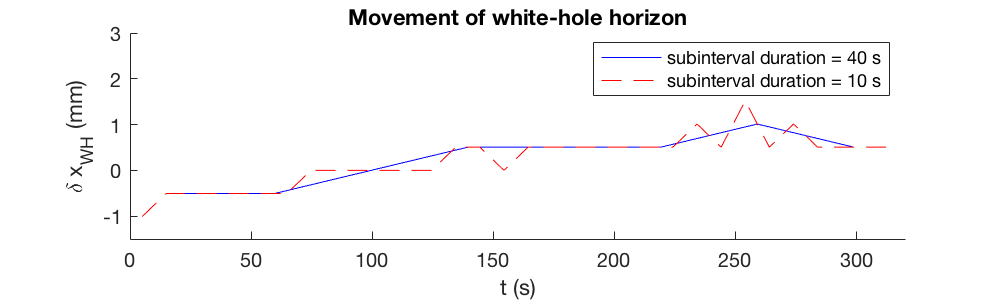} 
\caption{We show here a series of plots characterizing a flow which is gated at the downstream end. 
These are: 
(1) The obstacle height profile and the mean water height profile, averaged over the entire recording.  In vertical dashed lines are shown the positions of the black- and white-hole horizons.  
(2) The flow velocity $v(x)$ and wave speed $c(x)$, calculated using Eqs.~(\ref{eq:v_and_c}) in the Appendix.
(3) A zoom on the region shown by a dotted rectangle in panel (1), but now with the mean water height calculated separately in a series of $\sim 10 \, {\rm s}$ subintervals, all shown simultaneously.  The thickness of the curve thus indicates the degree of variation in time.  
(4) The position of the white-hole horizon, calculated in subintervals of duration $\sim 40 \, {\rm s}$ (in sold blue) and $\sim 10 \, {\rm s}$ (in dashed red).  
The flow rate is $Q=9.8\,{\rm L}/{\rm min}$, or (reducing to two dimensions) $q = Q/W = 3.1 \times 10^{-3} {\rm m}^{2}/{\rm s}$ (where $W = 5.3\,{\rm cm}$ is the channel width).
}
\label{fig:gate}
\end{figure}

In Fig.~\ref{fig:gate} is shown a series of plots associated with a gated flow, where the upper gate at the end of the water channel has been partially closed, leaving a gap of 0.8 cm at the bottom through which the flow can pass (the full downstream water height is 3.1 cm). 
The obstacle here has the same shape as that used in Figs.~\ref{fig:nogate}-\ref{fig:nogate_cslow}, but its dimensions are twice as large.
The plots are analogous to those shown in Fig.~\ref{fig:nogate}, showing the mean water depth averaged over all time, the mean water depth averaged over $\sim 10$ s subintervals, and the position of the white-hole horizon as a function of time. 
As before, there is a noticeable degree of variation of the background in the downstream region, while the upstream region is relatively stable.

\begin{figure}
\includegraphics[width=\columnwidth]{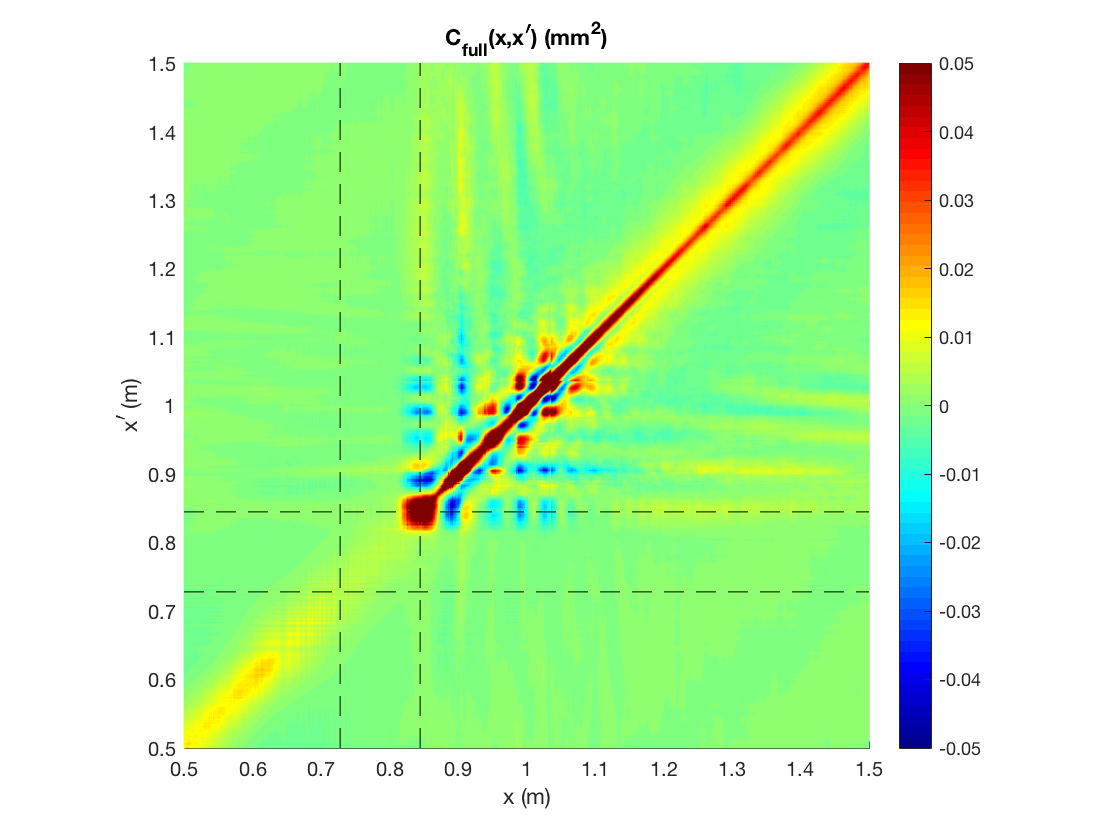}\\
\includegraphics[width=\columnwidth]{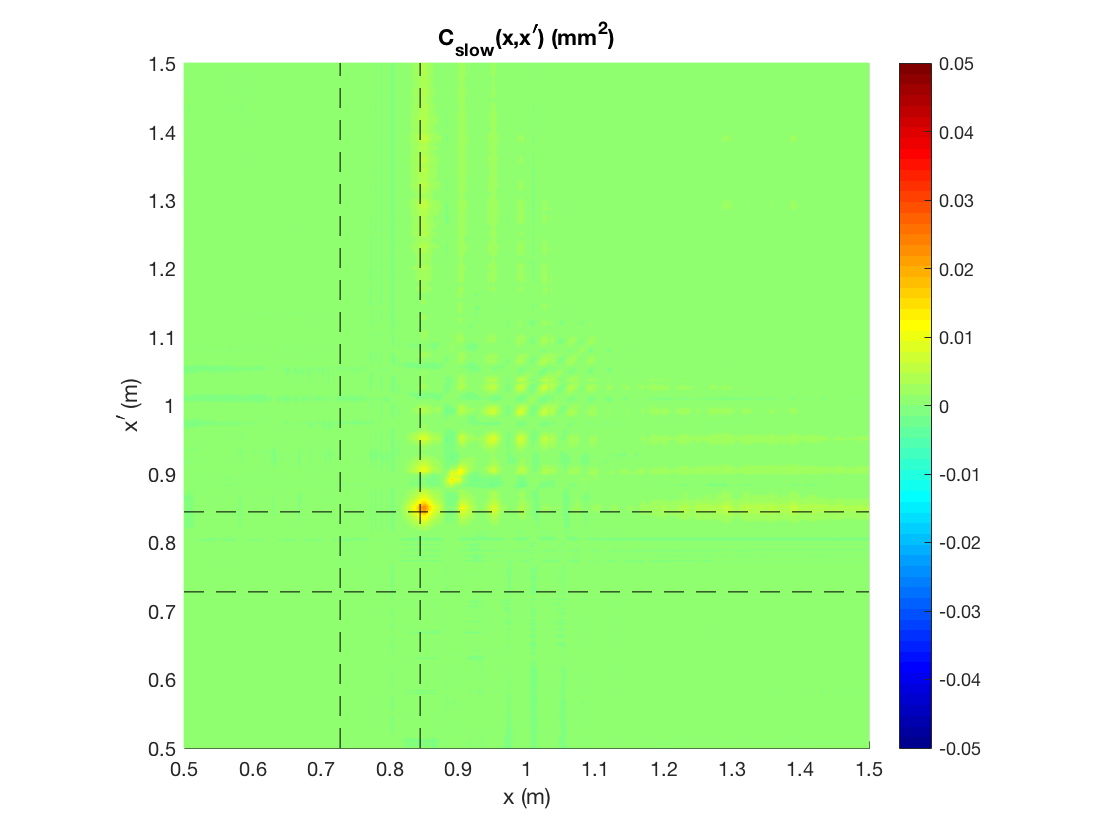}\\
\includegraphics[width=\columnwidth]{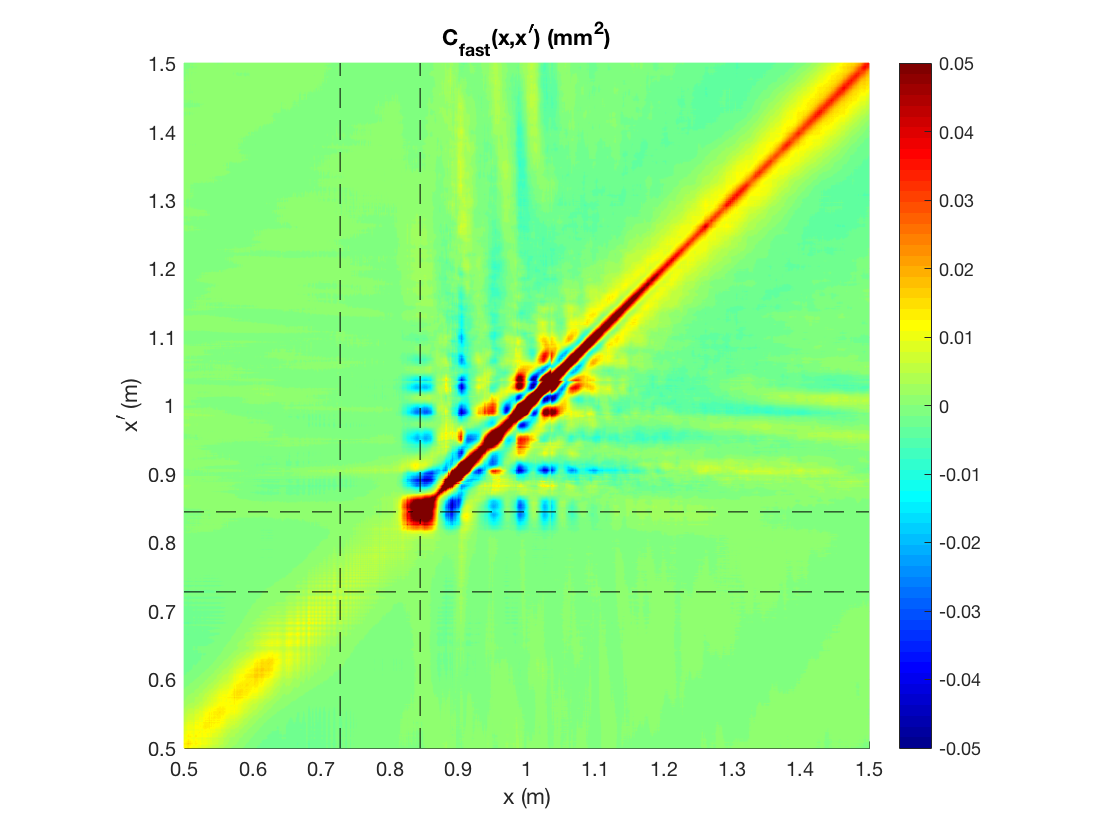}
\caption{The two-point correlation function of free surface fluctuations in the downstream region, for the same gated flow of Fig.~\ref{fig:gate}. 
The three correlation functions correspond to different definitions of the fluctuation $\delta h$: the full fluctuations (where the background is the time-average over the entire recording), the ``slow'' fluctuations associated with movement of the background (found by averaging over 10-second subintervals), and the ``fast'' fluctuations on top of this slowly-varying background.  
}
\label{fig:gate_checkerboard}
\end{figure}

A very noticeable difference with respect to Fig.~\ref{fig:nogate} is that, while there is still an undulation as the flow decelerates on the downstream side of the obstacle, it occurs on top of the obstacle, and not some distance downstream after an abrupt change in the flow.
The undular hydraulic jump is suppressed by the presence of the gate at the downstream end of the flume: there is a ``backwater effect''~\cite{ChansonBook} that prevents its appearance far downstream from the obstacle.  
Consequently, since the flow decelerates rather quickly, the supercritical region between the two horizons is rather short.
Less intuitively, the undulation itself is also significantly shorter than its counterpart in non-gated flows.~\footnote{This is likely related to the existence of a threshold flow velocity ($\sim 23\,{\rm cm}/{\rm s}$ in water) below which no zero-frequency solution for surface waves exists~\cite{Rousseaux-et-al-2010}; if this is crossed, then the undulation will be suppressed, and any corresponding checkerboard pattern will vanish.  In Figs.~\ref{fig:gate} and~\ref{fig:gate_checkerboard}, this happens at around $x = 1.1\,{\rm m}$. \label{fn:Landau}}
Numerical simulations of the flow are able to reproduce this observation (see the Supplemental Material).

\begin{figure}
\includegraphics[width=\columnwidth]{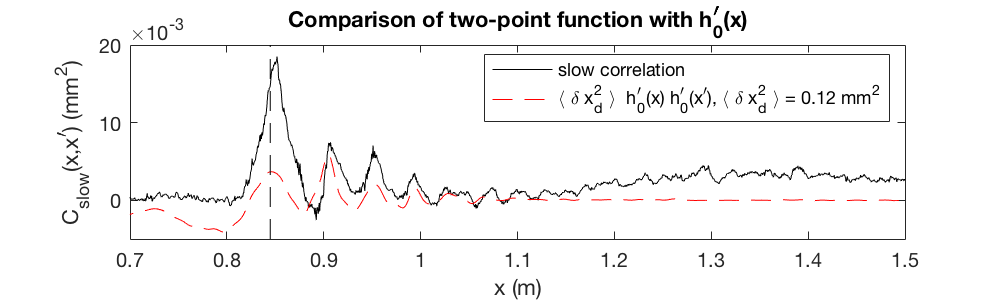} 
\caption{In black is shown a cross-section of the ``slow'' correlation of Fig.~\ref{fig:gate_checkerboard}, with $x^{\prime} \approx 0.85 \, {\rm m}$ fixed close to the point where $C_{\rm slow}(x,x^{\prime})$ reaches its maximum value.  In red is shown the spatial derivative of the mean water height $h_{0}(x)$ (averaged over the entire recording), multiplied by a prefactor found by fitting it to the black curve.  If the two curves were in good agreement, this would have been associated with a value of $\left\langle \delta x_{d}^{2} \right\rangle$ through prediction~(\ref{eq:corr_pred}); however, given the clear discrepancies between the two curves, prediction~(\ref{eq:corr_pred}) is not valid here and the quoted value of $\left\langle \delta x_{d}^{2} \right\rangle$ is not physically meaningful.
}
\label{fig:gate_cslow}
\end{figure}

With the undulation occurring on top of the obstacle, we might intuitively expect that it has no room to move around longitudinally, and that the checkerboard pattern observed in non-gated flows to be associated with the movement of the undular jump is strongly suppressed.  
This indeed seems to be the case, as illustrated by the two-point correlation functions for the gated flow, shown in Fig.~\ref{fig:gate_checkerboard}. 
As before, we show first the full correlation function, followed by its division into ``slow'' and ``fast'' contributions.
The obvious difference with respect to the previous case is that the ``slow'' contribution to the correlations is much less significant; indeed, the full correlation function is practically indistinguishable from the ``fast'' contribution alone.
Moreover, while the ``slow'' fluctuations generate a checkerboard pattern reminiscent of that in the non-gated flow (albeit much weaker), for the gated flow they
are not straightforwardly related to shifts in the longitudinal position of the undulation.  The most telling sign of this appears in Fig.~\ref{fig:gate_cslow}, where a cross-section of the ``slow'' correlation function is plotted alongside the spatial derivative of the water height (multiplied by a fitted prefactor) in order to check the validity of prediction~(\ref{eq:corr_pred}).  Unlike what was observed in Fig.~\ref{fig:nogate_cslow}, there is a clear discrepancy here, and the two profiles cannot be said to match.

Notice also that, in the ``slow'' contribution to the two-point correlation function, there is a clear offset in the far downstream region where the water surface is flat.  This indicates a significant {\it vertical} shift in the water height as opposed to a horizontal one, since horizontal motion of a flat surface generates no fluctuation.  We do not know the cause of this vertical motion, though it could be related to the same kind of back-reaction effects observed in~\cite{Patrick-et-al-2021}, due to the transport of mass by surface waves.  In any case, since the value of this offset is about an order of magnitude smaller than the amplitude of the oscillations seen on the right of Fig.~\ref{fig:nogate_cslow}, we cannot rule out its presence in Fig.~\ref{fig:nogate_cslow} and are thus 
unable to pin down this feature as being unique to gated flows.

Finally, we note that, while there is a clear but weak checkerboard pattern associated with ``slow'' fluctuations, there is a much stronger checkerboard-like pattern associated with ``fast'' fluctuations (in addition to the strong autocorrelation along the diagonal $x=x^{\prime}$).  Since we cannot link it straightforwardly to the motion of the background flow, we believe that this checkerboard-like feature is really due to the scattering of surface waves, in particular to a scattering process that efficiently produces short-wavelength dispersive modes just downstream of the white-hole horizon.  It is known in Analogue Gravity that a long-wavelength counter-current wave incident on the white-hole horizon from the downstream side will scatter into two short-wavelength dispersive modes, which play the role of the Hawking pairs~\cite{Mayoral-et-al-2011,Balbinot-Fabbri-Mayoral-2013}.  Fourier analysis of the noise in the far downstream region indicate that the required incident modes are indeed present (see the last section of the Supplemental Material).  Moreover, the presence of a checkerboard-like feature seems to imply that the two dispersive modes are correlated with each other; for, in the absence of such correlations, each plane wave mode $\propto e^{i k\left(\omega\right) x - i \omega t}$ would contribute to the two-point function via a term of the form $e^{i k\left(\omega\right)\left(x-x^{\prime}\right)}$ that is a function of $x-x^{\prime}$ only.  (In the Supplemental Material, we also see signs of these correlations in Fourier space.)  It is thus tempting to conclude that the rather complex correlation pattern in the first and third panels of Fig.~\ref{fig:gate_checkerboard} is to some degree a signal of the classical analogue of the Hawking process taking place at the white-hole horizon.
This could be verified by a careful analysis of the corresponding scattering coefficients, which is beyond the scope of this paper but is an intriguing direction for further research.

\section{Summary and discussion}
\label{sec:Discussion}

We have performed a series of experimental runs in which a stationary transcritical flow is realised in a narrow 1D channel.  
The analogue white-hole in the downstream region (where the flow passes from super- to subcritical) is observed to coincide with the onset of a visible motion of the background, as defined by averaging over subintervals much shorter than the full duration of the recording.
Broadly speaking, the flows can be categorized into two groups according to the behavior of the white-hole horizon and the associated undulation, which is linked to the presence or absence of a gate restricting the upper part of the flow at the downstream end of the channel.  In the absence of such a gate, the white-hole horizon occurs some distance downstream from the obstacle, being generated spontaneously by the occurrence of an undular hydraulic jump.  In the two-point function of free surface deformations, we observe a checkerboard pattern in this downstream region that is associated with slow fluctuations, and can be reasonably well-described by a degree of randomness in the longitudinal position of the jump.  On the other hand, when a gate is present, the white-hole horizon and its associated undulation tend to occur on top of the obstacle.  In this case, while there is still a checkerboard-like pattern in the vicinity of the undulation, it is not associated with slow fluctuations nor with a straightforward longitudinal shift of the undulation.

Although we are not yet able to give precise physical reasons for the observed behavior, it seems to be physically intuitive, at least in part.  That the undulation is longitudinally mobile in the free flow seems reasonable, given that the flow is not restricted at the downstream end, nor much at the white-hole horizon which occurs some distance from the obstacle.  Indeed, being far from the obstacle, the undulation in this case occurs in a background which is approximately translation-invariant, so it makes sense that there is some translational freedom that manifests through a degree of variability in the longitudinal position. 

By contrast, the ``attachment'' of the undulation to the obstacle in the case of gated flows removes this translational freedom, and the checkerboard pattern associated with slow fluctuations is strongly suppressed.
The background motion is no longer longitudinal, seeming to be largely in the vertical direction.
Nevertheless, there remains a checkerboard-like pattern associated with fast fluctuations.  
As pointed out above (and elaborated upon in the Supplemental Material), there are strong indications that a stimulated analogue Hawking process is occurring -- namely, the required ancestor modes are clearly present on the downstream side, and the two dispersive modes that ought to be produced by the Hawking process are correlated.
It is thus tempting to view the correlation pattern in this case as a signature of the Hawking effect.  
A more precise and convincing proof would be to extract the scattering coefficients associated with this process; these should be 
approximately given by a Planck spectrum~\footnote{We should recall that, at low frequencies, the thermal behavior $\sim \sqrt{\omega_{H}/\omega}$ of the associated scattering coefficients is recovered only in a finite range $\left[\omega_{c}\,,\,\omega_{H}\right]$, where $\omega_{H}$ is the Hawking temperature and $\omega_{c}$ is a low-frequency cutoff associated with the finite length of the supercritical region~\cite{Robertson-Michel-Parentani-2016}.  The shortness of the supercritical region compared with the non-gated case may help to explain the relative absence of a strong signal for ``slow'' frequencies.  We estimate $\omega_{H} \sim 1\,{\rm Hz}$, so there is room for a Hawking-type signal in the ``fast'' frequency range $\omega > 1/\left(10\,{\rm s}\right)$.} whose temperature is determined by the flow properties at the horizon~\cite{Mayoral-et-al-2011,Balbinot-Fabbri-Mayoral-2013}.
With the particular flow in question, this is difficult to achieve, as the produced dispersive modes exist only in a relatively short region downstream from the white-hole horizon, where the background flow is varying a lot (and is not well-known thanks to the likely presence of flow circulation on the bottom of the channel -- see ``Numerical simulation of background flow'' in the Supplemental Material).  Therefore, such an analysis likely requires the realisation of a gated flow where the undulation and the dispersive modes exist over an extended region where the flow is relatively flat.

Finally, it should be pointed out that the precise origins of the movement of the undulation in the free flows are not known, despite the possibility of such movement being rather intuitive.  In particular, 
we cannot rule out the possibility that these slow fluctuations are themselves the result of an analogue Hawking effect.  For the Hawking process, being described by a thermal spectrum, is (in 1D) most prevalent for very low frequencies. 
Moreover, since the downstream side of the flow lies far from the obstacle in a regime where the system is almost translation-invariant, it makes sense that one of the zero-frequency solutions be characteristic of this symmetry, and thus that $\delta h \propto h_{0}^{\prime}(x)$ be a solution of the linearized wave equation on top of such a flow.  A more definitive answer to this question requires a more precise treatment of the linearized wave equation on top of a background flow where the undulation is already at its saturated level.  This is beyond the scope of the present work, but would be an interesting direction for further research.

\section*{Acknowledgments}
We would like to thank Romain Bellanger, Brice Bechet and Jean-Marc Mougenot for their help concerning the experimental aspects of this work.
This work pertains to the French government program ``Investissements d'Avenir'' (EUR INTREE, reference ANR-18-EURE-0010, and LABEX INTERACTIFS, reference ANR-11-LABX-0017-01).
It benefited from the support of the project OFHYS of the CNRS 80 Pprime initiative in 2019-2020 and from an internal funding scheme ACI Pprime.
AF acknowledges partial financial support by the Spanish grants from Ministerio de Ciencia e Innovaci\'{o}n FIS2017-84440-C2-1-P funded by MCIN/AEI/10.13039/501100011033 ``ERDF A way of making Europe'', Grant PID2020-116567GB-C21 funded by MCIN/AEI/10.13039/501100011033, and the project PROMETEO/2020/079 (Generalitat Valenciana).
SR and AF acknowledge support from the French National Research Agency through grant no. ANR-20-CE47-0001 associated with the project COSQUA (Cosmology and Quantum Simulation).

\bibliography{bibliography}

\newpage

\section*{Supplemental Material}

\subsection*{Background and fluctuations}

In this section, we give some theoretical details concerning the decomposition of the full water height $h(x,t)$ into a ``background'' and a fluctuation propagating on top of that background.

There is inevitably some ambiguity in the decomposition of the full water depth into a background plus a fluctuation:
\begin{equation}
h(x,t) = h_{0}(x,t) + \delta h(x,t) \,.
\end{equation}
Only the full water depth, $h(x,t)$, is measured experimentally.  The decomposition into the background, $h_{0}(x,t)$, and the fluctuation, $\delta h(x,t)$, is imposed in the data analysis, and is inherently ambiguous, for there are many ways to define $h_{0}(x,t)$ (at least when averaging over time, rather than taking an ensemble average).  The simplest way is simply to define a time-independent $h_{0}(x)$ as being the average of $h(x,t)$ over the entire duration of the recording.  If the background profile drifts over this duration, then that drift will be included in the fluctuation $\delta h(x,t)$.  We may, however, split the full duration into shorter subintervals, defining $h_{0}(x,t)$ as the local average over each segment.  Crucially, we should see a clear dependence on the duration of the segments if there is a separation of scales between the typical time associated with the passage of surface waves and the typical time associated with the evolution of the background.  If the segment duration is much shorter than the latter, then the background will be approximately constant over each segment, and the movement of the background will be included in $h_{0}(x,t)$ rather than in $\delta h(x,t)$, the latter then capturing the ``true'' surface waves.  On the other hand, if the segment duration is large compared to the time scale associated with the movement of the background, then this movement will be included in $\delta h(x,t)$ and will be treated as a fluctuation on an equal footing with the surface waves.

This, of course, affects all quantities used to characterise the statistical properties of the fluctuations.  In particular, the (time-averaged) two-point correlation function $\left\langle \delta h\left(x,t\right) \, \delta h\left(x^{\prime},t^{\prime}\right) \right\rangle_{t}$ is affected.  To see this explicitly, let us write the fully time-dependent water depth as:
\begin{equation}
h(x,t) = h_{0}(x) + \delta h_{\rm slow}(x,t) + \delta h_{\rm fast}(x,t) \,.
\end{equation}
We identify $h_{0}(x) + \delta h_{\rm slow}(x,t)$ with the time-dependent background, $h_{0}(x,t)$.  $h_{0}(x)$ is just the average of $h(x,t)$ over the entire duration of the recording.  The division of the remainder into $\delta h_{\rm slow}(x,t)$ and $\delta h_{\rm fast}(x,t)$ is then a matter of choice, depending on the duration of the sub-intervals averaged over in order to define $h_{0}(x,t)$.
The equal-time two-point function for the total height is
\begin{multline}
\left\langle h(x,t) h(x^{\prime},t) \right\rangle_{t} = h_{0}(x) h_{0}(x^{\prime}) + \left\langle \delta h_{\rm slow}(x,t) \delta h_{\rm slow}(x^{\prime}t) \right\rangle_{t} \\ + \left\langle \delta h_{\rm fast}(x,t) \delta h_{\rm fast}(x^{\prime},t) \right\rangle_{t} \,.
\end{multline}
The cross-terms must vanish because $h_{0}(x)$ and $\delta h_{\rm slow}(x,t)$ are themselves defined as averages over time, with the average of the perturbation over each sub-interval necessarily vanishing.  Now, $\left\langle h(x,t) h(x^{\prime},t) \right\rangle_{t}$ and $h_{0}(x) h_{0}(x^{\prime})$ are unambiguously defined for any single recording.  The only ambiguity rests in the division into the two-point functions for $\delta h_{\rm slow}$ and $\delta h_{\rm fast}$.  So we cannot lose any information when exploring different partitions of the whole duration into sub-intervals; we only divide the information differently into ``slow'' and ``fast'' perturbations.

\subsection*{Extraction of relevant quantities}

The mean water height $h_{0}(x)$ is calculated by averaging $h(x,t)$ over all time, with the typical recording duration being 340 seconds.  Once this is known, the (depth-averaged) flow velocity profile $v(x)$ and the wave speed $c(x)$ are calculated using
\begin{equation}
v(x) = \frac{q}{h_{0}(x)} \,, \qquad c(x) = \sqrt{g \, h_{0}(x)} \,,
\label{eq:v_and_c}
\end{equation}
where $q = Q/b$ is the flow rate per unit width ($b = 5.3 \, {\rm cm}$ is the width of the channel) and $g = 9.8 \,{\rm m}/{\rm s}^{2}$ is the acceleration due to gravity.  These expressions hold in the absence of significant vorticity and when the surface is relatively flat $(i.e., \left|{\rm d}z/{\rm d}x\right| \ll 1)$.  They are expected to be accurate in the upstream region, though the presence of vorticity downstream from the obstacle is expected to make them less accurate there.  The horizons are simply defined as the points where $v(x) = c(x)$, the black-hole horizon in the upstream region where the flow passes from sub- to supercritical, and the white-hole horizon in the downstream region where it returns to subcritical.  It is interesting to note that, even though the white-hole horizon occurs in the downstream region where the validity of Eqs.~(\ref{eq:v_and_c}) could be questioned, it typically coincides well with the beginning of the region where significant movement of the background is observed.

The instantaneous free surface deformation is simply defined as $\delta h(x,t) = h(x,t) - h_{0}(x)$.
The full two-point correlation function is then just the average over all time of the product of $\delta h(x,t)$ observed simultaneously at two points:
\begin{align}
C_{\rm full}(x,x^{\prime}) &= \left\langle \delta h(x,t) \, \delta h(x^{\prime},t) \right\rangle_{t} \nonumber \\
&= \frac{1}{N_{t}} \sum_{j=1}^{N_{t}} \delta h(x,t_{j}) \, \delta h(x^{\prime},t_{j}) \,,
\end{align}
where $N_{t}$ is the total number of discrete measurement times (typically $N_{t} = 2^{13} = 8192$).

The division into subintervals is achieved by factorizing $N_{t} = n_{\rm sub} \times N_{t}^{\rm sub}$, representing a total of $n_{\rm sub}$ subintervals each containing $N_{t}^{\rm sub}$ discrete measurement times.  Then profile of the mean water height $h_{0}^{\rm sub}(x,t)$ can be calculated within each subinterval.  The ``slow'' fluctuation is defined as the difference between this and the overall mean $h_{0}(x)$, while the ``fast'' fluctuation is defined as the remainder:
\begin{align}
\delta h_{\rm slow}(x,t) &= h_{0}^{\rm sub}(x,t) - h_{0}(x) \nonumber \\
\delta h_{\rm fast}(x,t) &= \delta h(x,t) - \delta h_{\rm slow}(x,t) = h(x,t) - h_{0}^{\rm sub}(x,t) \,.
\end{align}
The ``slow'' and ``fast'' contributions to the two-point correlation function are then straightforwardly defined:
\begin{align}
C_{\rm slow}(x,x^{\prime}) &= \left\langle \delta h_{\rm slow}(x,t) \, \delta h_{\rm slow}(x^{\prime},t) \right\rangle_{t} \,, \nonumber \\
C_{\rm fast}(x,x^{\prime}) &= \left\langle \delta h_{\rm fast}(x,t) \, \delta h_{\rm fast}(x^{\prime},t) \right\rangle_{t} \,.
\end{align}

The derivative $h_{0}^{\prime}(x)$, needed to compare the observed ``slow'' two-point correlation function with the prediction~(\ref{eq:corr_pred}), is calculated as follows.  The height profile $h_{0}(x)$ is smoothed by applying a window in the Fourier transform so as to remove noise of high spatial frequency.  The window takes the form
\begin{equation}
W(k) = \frac{1}{2}\left( {\rm tanh}\left( \frac{k+k_{\rm cut}}{\sigma_{k}} \right) - {\rm tanh}\left( \frac{k-k_{\rm cut}}{\sigma_{k}} \right) \right) \,,
\end{equation}
where we take $k_{\rm cut} = 400 \, {\rm m}^{-1}$ and $\sigma_{k} = 50 \, {\rm m}^{-1}$.  We then multiply by $i\,k$ and take the inverse Fourier transform, which yields $h_{0}^{\prime}(x)$ for the relevant values of $x$.  (It generates rather large deviations at the edges of the spatial window, but these are not relevant for our purposes.)

In comparing the two-point function with the prediction~(\ref{eq:corr_pred}), we fix $x^{\prime}$ to be close to the first maximum of $C_{\rm slow}(x,x^{\prime})$, and then simply perform a linear fit of the amplitude needed to match $h_{0}^{\prime}(x) \, h_{0}^{\prime}(x^{\prime})$ with the observed two-point function.  We only include those points within the displayed spatial window (corresponding to the dotted rectangle on the plots of the mean water height).

\subsection*{Numerical simulation of background flow}

An original two-dimensional free-surface flow code was used to simulate numerically the transcritical flows studied experimentally in this paper. A projection method is applied to the incompressible variable density Navier-Stokes equations to decouple velocity and pressure unknowns. Away from the interfaces (water-air and obstacle-water), partial differential operators (divergence, gradient, Laplacian operator) and nonlinear terms are discretized on a fixed Cartesian grid using standard second-order finite difference approximations. Several techniques are used to account for the presence of the two interfaces while avoiding the generation of conformal meshes. An Immersed Boundary Method~\cite{Bouchon-Dubois-James-2012} enforces the no-slip boundary condition on the rigid obstacle and the free surface evolution is tackled with the Level-Set technique~\cite{Osher-Sethian-1988}.

\begin{figure}
\includegraphics[width=\columnwidth]{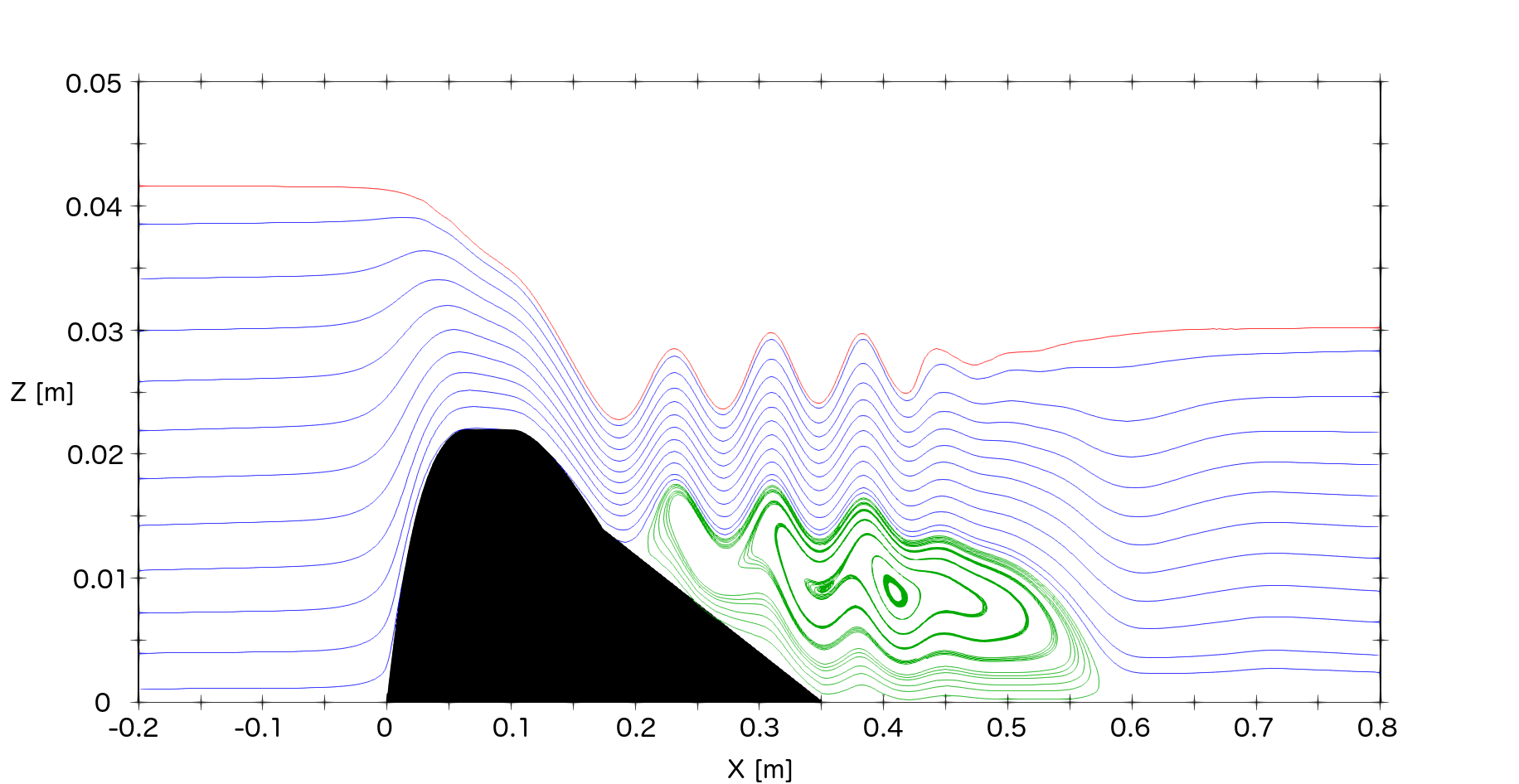}
\caption{Numerical simulation of background flow.  Here are shown the streamlines of the flow over the obstacle as produced by the numerical simulation (see details in text), for a flow with a gate at the downstream end.  The short undulation on top of the obstacle is clearly seen.  It is also observed to be associated with a circulation layer on the bottom of the flow, shown here in green.  This makes the upper part of the flow effectively flat, allowing the undulation to extend further downstream than it otherwise would have.}
\label{fig:simulations}
\end{figure}

Results for a gated flow, with the obstacle of Figs.~\ref{fig:gate}-\ref{fig:gate_cslow}, are given in Figure~\ref{fig:simulations}.  This shows the streamlines of the flow once a steady state has been reached, starting from an initial water height of 3.3 cm (so that the obstacle is always completely submerged).  The simulation neglects the transverse direction, so in order to mimic the increased effective dissipation due to the narrowness of the channel and friction at the walls, the viscosity of water has been increased by a factor of 10.  The results are in qualitative agreement with what is seen experimentally in Fig.~\ref{fig:gate}: after an initial acceleration on top of the obstacle, the flow immediately decelerates and induces an undulation, which exists over a short region and vanishes in the asymptotic downstream region.

Notice that the undulation appears to be coupled with a circulation layer on the bottom of the channel, and that the non-circulatory layer on top has a depth which is essentially constant in the vicinity of the undulation, effectively cancelling out the downstream slope of the obstacle.  This helps to explain why the undulation extends as far as it does and why it disappears rather abruptly.  The undulation is essentially a (rather large-amplitude) zero-frequency free surface deformation, satisfying $\omega(k)=0$ where $k$ is the wave vector of the undulation.  This means that the phase velocity of the undulation, $\omega/k$, vanishes, but by addition of velocities this is just equal to $v_{\rm ph}(k) + v_{\rm flow}$, where $v_{\rm ph}(k)$ is the phase velocity in the rest frame of the fluid and $v_{\rm flow}$ is the flow speed.  Due to capillary effects, $v_{\rm ph}(k)$ has a minimum at $\sim 23 \, {\rm cm}/{\rm s}$ (this is the threshold mentioned in footnote~\ref{fn:Landau}), so $v_{\rm flow}$ needs to be larger than this in order for a non-trivial zero-frequency solution to exist.  A straightforward application of the formula $\bar{v}(x) = q/h(x)$ (for the depth-averaged flow velocity $\bar{v}$) indicates that $\bar{v}(x)$ dips below $23\,{\rm cm}/{\rm s}$ significantly before the end of the undulation.  However, in the numerical results of Fig.~\ref{fig:simulations} we see that the non-circulating flow near the surface has a relatively shallow constant depth some distance from the obstacle, so that the surface flow dips below $23\,{\rm cm}/{\rm s}$ only towards the end of the circulation layer.  We indeed see that the undulation ends in rough coincidence with the end of the circulation layer.

\subsection*{Dependence on subinterval duration}

\begin{figure}
\includegraphics[width=\columnwidth]{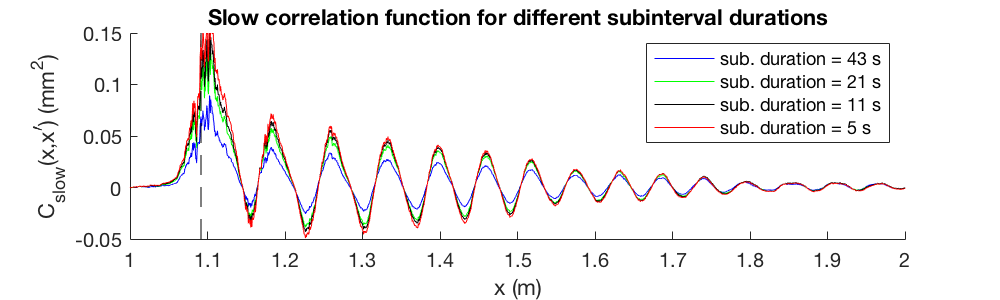} \\
\includegraphics[width=\columnwidth]{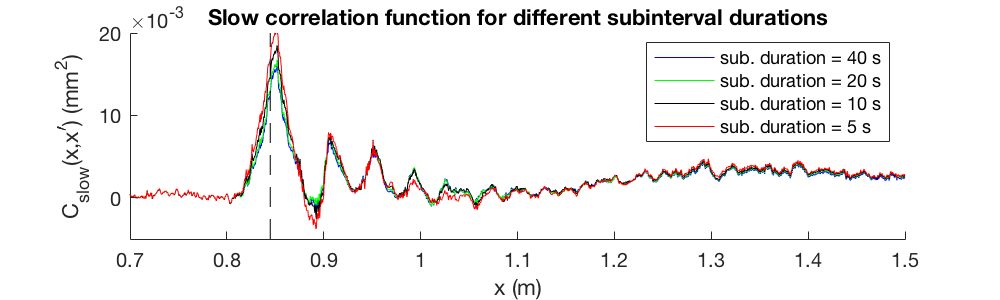}
\caption{$C_{\rm slow}(x,x^{\prime})$ of Figs.~\ref{fig:nogate} (upper, non-gated flow) and~\ref{fig:gate} (lower, for gated flow), for different values of the subinterval duration.  In both cases the profile is quite stable.}
\label{fig:ChangingSubDuration}
\end{figure}

In Fig.~\ref{fig:ChangingSubDuration} are plotted the ``slow'' correlation functions of Figs.~\ref{fig:nogate_cslow} and~\ref{fig:gate_cslow} for varying durations of subintervals in which the background is defined by time-averaging.  What is most notable is the relative stability of the curves with respect to the subinterval duration.  We do see a significant change in the non-gated case when changing from a duration of 43 seconds to 21 seconds.  This indicates that we cross the time scale associated with the slow fluctuations: a duration of 43 seconds is long enough that the background has already undergone a noticeable change, and this change gets labelled as ``fast'' rather than ``slow''.

\subsection*{Correlations in Fourier space
\label{sec:correlations_fourier}}

\begin{figure}
\includegraphics[width=0.9\columnwidth]{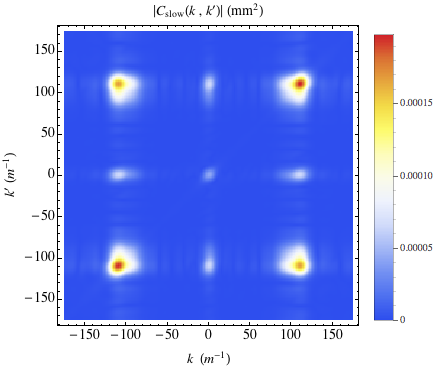}
\caption{Magnitude of the two-point correlation function in Fourier space, for the ``slow'' fluctuations in the downstream region of the non-gated flow of Figs.~\ref{fig:nogate}-\ref{fig:nogate_cslow}.}
\label{fig:corr_Fourier}
\end{figure}

Given that scattering on a stationary background proceeds independently for different frequencies, it can be illuminating to consider correlations between different Fourier modes rather than between different points in space.  To this end, restricting our attention to the non-gated flow of Fig.~\ref{fig:nogate}, we define the Fourier amplitudes
\begin{equation}
\delta h_{j}(k,t) = \frac{1}{x_{2}-x_{1}} \int_{x_{1}}^{x_{2}} {\rm d}x \, H(x) \, e^{-i k x} \, \delta h_{j}(x,t) \,.
\label{eq:Fourier_defn}
\end{equation}
where $H(x)$ is a window function called the Hamming window:
\begin{equation}
H(x) = \frac{25}{46} + \frac{21}{46} \, {\rm cos}\left( 2\pi \frac{x-\frac{1}{2}\left(x_{1}+x_{2}\right)}{x_{2}-x_{1}}\right) \,.
\end{equation}
This choice of window is convenient as it suppresses the first side-lobe in the Fourier transform over a finite window size~\cite{Euve-et-al-2016}.
We do not restrict $k$ to discrete values; this means that while Fourier amplitudes which are nearby in $k$ are not independent, they are smooth rather than pixellated.
The subscript $j$ stands for the various definitions of $\delta h(x,t)$ defined above, with $j \in \left\{ {\rm full} \,,\, {\rm slow} \,,\, {\rm fast} \right\}$.
Then we can define the Fourier-space two-point correlation function analogously to that in position-space:
\begin{equation}
C_{j}(k,k^{\prime}) = \left\langle \delta h_{j}(k,t) \, \delta h_{j}^{\star}(k^{\prime},t) \right\rangle_{t} \,.
\end{equation}
By construction, we have $C_{j}(k,k^{\prime}) = C_{j}^{\star}(k^{\prime},k)$; moreover, since $\delta h(x,t)$ is real and consequently $\delta h(k,t) = \delta h^{\star}(-k,t)$, we also have $C_{j}(k,k^{\prime}) = C_{j}(-k^{\prime},-k)$.
These mean that the magnitude $\left|C_{j}(k,k^{\prime})\right|$ must be symmetric under reflection in both the diagonal $k=k^{\prime}$ and the anti-diagonal $k=-k^{\prime}$.

\subsubsection*{Non-gated flow}

In Fig.~\ref{fig:corr_Fourier} is shown the magnitude of $C_{\rm slow}(k,k^{\prime})$ for the Fourier-space correlation function in the downstream region ({\it i.e.}, in the vicinity of the undulation) of the non-gated flow of Fig.~\ref{fig:nogate}.    
Here, we have chosen $x_{1} = 1.15\,{\rm m}$ and $x_{2} = 2.06\,{\rm m}$, so as to capture the entire downstream region beyond the hydraulic jump.
The key observation here is the simple $3 \times 3$ structure of the correlations, occurring between three distinct wave vectors: $k=0$ and $k=\pm k_{u}$, where $k_{u}$ is the wave vector associated with the undulation.  (We see that $k_{u} \sim 110\,{\rm m}^{-1}$, corresponding to a wavelength of $2\pi/k_{u} \sim 6\,{\rm cm}$, which is consistent with the undulation pattern seen in Figs.~\ref{fig:nogate}-\ref{fig:nogate_cslow}.) 
This is consistent with the physical interpretation given in Eq.~(\ref{eq:corr_pred}). 
Fourier transforming $\delta h(t,x) \approx -\delta x_{d}(t) \, h_{0}^{\prime}(x)$ in space, we have $\delta h(k,t) \approx - \delta x_{d}(t) \, \left[ h_{0}^{\prime} \right]_{k}$, where $\left[ h_{0}^{\prime} \right]_{k}$ is just the Fourier transform of $h_{0}^{\prime}(x)$.  
Then 
\begin{equation}
C_{\rm slow}(k,k^{\prime}) \approx \left\langle \delta x_{d}^{2}(t) \right\rangle_{t} \, \left[ h_{0}^{\prime} \right]_{k} \, \left[ h_{0}^{\prime} \right]_{k^{\prime}} ^{\star} \,.
\end{equation}
This particular form of $C_{\rm slow}(k,k^{\prime})$ requires that, not only are the required symmetries described above respected, but we also have $\left|C_{\rm slow}(k,k^{\prime})\right| = \left|C_{\rm slow}(k,-k^{\prime})\right|$ (and similarly for $k \to -k$).  That is, the magnitude of the correlation function is symmetric under reflection in the horizontal and vertical axes.  Although this is only an approximate symmetry that depends on the validity of prediction~(\ref{eq:corr_pred}), it is borne out by the correlations in Fig.~\ref{fig:corr_Fourier}: the ratio $\left|C_{\rm slow}(k_{u},-k_{u})\right|/\left|C_{\rm slow}(k_{u},k_{u})\right|$ is $0.93$ where $\left|C_{\rm slow}(k_{u},k_{u})\right|$ reaches its maximum value at $k_{u} \sim 110\,{\rm m}^{-1}$.

\begin{figure}
\includegraphics[width=0.9\columnwidth]{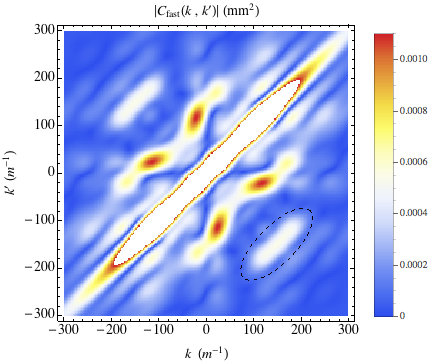}
\caption{Magnitude of the two-point correlation function in Fourier space, for the ``fast'' fluctuations in the vicinity of the undulation in the gated flow of Figs.~\ref{fig:gate}-\ref{fig:gate_cslow}.
The dashed ellipse points out the feature of interest: a correlation between the short-wavelength dispersive modes of positive and negative wave number.}
\label{fig:corr_Fourier_gate}
\end{figure}

\subsubsection*{Gated flow}

In Fig.~\ref{fig:corr_Fourier_gate} we show the magnitude of the Fourier-space correlation function $C_{\rm fast}(k,k^{\prime})$ for the region on top of the obstacle in the gated flow of Figs.~\ref{fig:gate}-\ref{fig:gate_cslow},
from $x_{1} = 0.8\,{\rm m}$ to $x_{2} = 1.1\,{\rm m}$.  
(We choose to show this for the ``fast'' fluctuations since it is these that generate the main checkerboard-like feature in the position-space correlation functions of Fig.~\ref{fig:gate_checkerboard}.)
Since the flow varies a lot in this region, the normal modes are not strict plane waves and there will be considerable ``smudging'' of the correlation pattern.  Nevertheless, it is sufficient to show the main feature of interest:
a clear correlation between positive and negative $k$, extending from about $\left|k\right|=100\,{\rm m}^{-1}$ to $\left|k\right|=200\,{\rm m}^{-1}$
(and indicated by the dashed ellipse in Fig.~\ref{fig:corr_Fourier_gate}).  
This correlation between short-wavelength dispersive modes of positive and negative $k$ is precisely what is engendered by the Hawking scattering process~\cite{Mayoral-et-al-2011,Balbinot-Fabbri-Mayoral-2013}.  
More precisely, the dispersion relation around the positive zero-frequency solution $k_{0}$ can be written, to lowest order in a Taylor expansion, as $\omega \approx v_{g,0} \left( k - k_{0} \right)$, where $k_{0}$ is the zero-frequency mode and $v_{g,0}$ its group velocity.  Around the negative solution $-k_{0}$, it instead takes the form $\omega \approx v_{g,0} \left(k + k_{0}\right)$.  So, at fixed $\omega$ not too far from zero, the two short-wavelength solutions of the dispersion relation are $k_{\rm pos} = k_{0} + \omega/v_{g,0}$ and $k_{\rm neg} = -k_{0} + \omega/v_{g,0}$.  Those with the same $\omega$ are correlated by the scattering and thus correlated.  On the $\left(k_{\rm pos},k_{\rm neg}\right)$ plane, the locus of these correlations describes a line of slope $1$ centered at $\left(k_{0},-k_{0}\right)$; on the full $\left(k,k^{\prime}\right)$ plane, this line appears in both the lower right quadrant and the upper left quadrant.  This is exactly what we see in Fig.~\ref{fig:corr_Fourier_gate}, and we may read off $k_{0} \approx \left(150 \pm 10\right) \, {\rm m}^{-1}$ (corresponding to a wavelength $\lambda_{0} \approx \left(4.2 \pm 0.3\right) {\rm cm}$, compatible with the undulation seen in Fig.~\ref{fig:gate}).  
Unfortunately, we are unable to check this value against theoretical predictions due to the relatively large variation of the water height $h$ and the corresponding flow velocity $v=q/h$ in the region of interest.

\begin{figure}
\includegraphics[width=0.9\columnwidth]{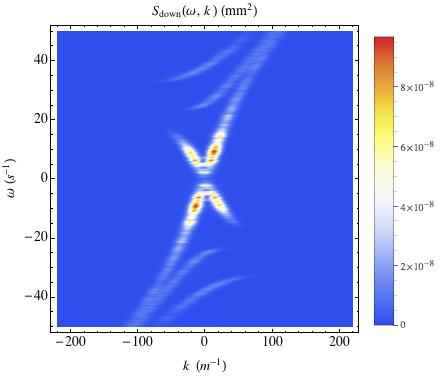} 
\includegraphics[width=0.9\columnwidth]{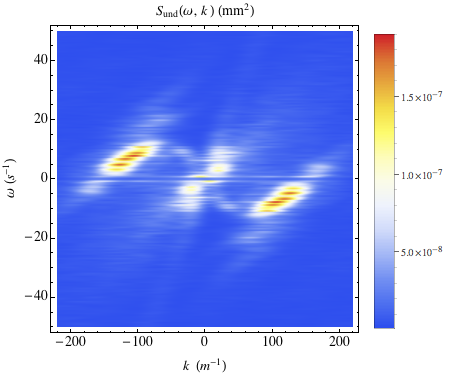}
\caption{Power spectra of fluctuations on the gated flow of Fig.~\ref{fig:gate}, both in the relatively flat downstream region(top) and in the vicinity of the undulation (bottom).}
\label{fig:gate_disp-rel}
\end{figure}

To further corroborate our claim that the correlations mentioned above are likely due to the analogue Hawking process associated with scattering at the white-hole horizon, we show in Fig.~\ref{fig:gate_disp-rel} the power spectra of the ``fast'' fluctuations in both the downstream region (where the background is relatively flat) and in the region of the undulation (where it varies a lot).  This power spectrum is defined somewhat analogously to Eq.~(\ref{eq:Fourier_defn}) for the Fourier amplitudes: we define $S(\omega,k) = \left\langle \left| \widetilde{\delta h}(\omega,k) \right|^{2} \right\rangle$, where
\begin{multline}
\widetilde{\delta h}(\omega,k) = \frac{1}{t_{2}-t_{1}} \frac{1}{x_{2}-x_{1}} \int_{t_{1}}^{t_{2}} dt \int_{x_{1}}^{x_{2}} dx \\ H(x) \, e^{i \omega t - i k x} \delta h(t,x) \,.
\end{multline}
The double Fourier transform $\widetilde{\delta h}(\omega,k)$ is calculated separately in each of 32 subintervals of $\sim 10\,{\rm s}$ each, and its squared magnitude is averaged over the subintervals to get $S(\omega,k)$.  This is particularly illuminating as the occupied modes will lie along the dispersion relation characterising the surface waves~\cite{Weinfurtner-et-al-2011}.   The key observations of Fig.~\ref{fig:gate_disp-rel} are: 
\begin{enumerate}
\item The counter-propagating branch of surface waves propagating in the upstream direction is significantly populated in the downstream region (probably because some of the co-propagating waves are reflected by the gate).  These will therefore be incident on the white-hole horizon, where they are expected to scatter into the two available dispersive modes by a process analogous to the Hawking effect~\cite{Mayoral-et-al-2011,Balbinot-Fabbri-Mayoral-2013}.
\item The dispersive branch of the dispersion relation in the vicinity of the undulation is clearly populated, most significantly on the {\it positive-energy} side of that branch.  This is important because the counter-propagating mode incident from the downstream side has positive energy, so it is expected to scatter preferentially into the positive-energy mode.
\end{enumerate}

\end{document}